\documentstyle[epsfig]{mn}

\title[$ H_0$ from Abell 773] {A measurement of $ H_0$ from Ryle
      Telescope, {\sl ASCA} and {\sl ROSAT} observations of Abell 773}
      \author[R. Saunders et al.]  {Richard Saunders,$^{1}$ R\"udiger
      Kneissl,$^{1}$ Keith Grainge,$^{1}$ William F. Grainger,$^{1}$
      \cr Michael E. Jones,$^{1}$ Alessia Maggi,$^{2}$ Rhiju Das,$^{1
      \star}$ Alastair C. Edge,$^{3}$ \cr Anthony N. Lasenby,$^{1}$
      G.G.  Pooley,$^{1}$, Shigeru J. Miyoshi,$^{4}$ Taisuke
      Tsuruta,$^{4}$ \cr Koujun Yamashita,$^{5}$ Yuzuru Tawara,$^{5}$
      Akihiro Furuzawa,$^{5}$ Akihiro Harada,$^{5}$ \cr Izamu
      Hatsukade$^{6}$ \\ $^{1}$Astrophysics Group, Cavendish Laboratory,
      Madingley Road, Cambridge CB3 0HE, UK \\ $^{2}$St. John's
      College, Cambridge, CB2 1TP, UK\\$^{3}$Department of Physics,
      South Road, Durham DH1 3LE, UK \\ $^{4}$Department of Physics,
      Kyoto Sangyo University, Kamigamo-Motoyama, Kito-ku, Kyoto 603,
      Japan \\ $^{5}$Department of Astrophysics, Faculty of Science,
      Nagoya University, Chikusa-ku, Nagoya 464-01, Japan \\
      $^{6}$Faculty of Engineering, Miyazaki University, 1-1
      Gakuen-kibanadai-nishi, Miyazaki 889-21,
      Japan\\$^{\star}$Present address: Physics Department, Stanford
      University, CA 94305-4060, USA \\} \date{}

\begin{document}

\maketitle

 \begin{abstract} We present new Ryle Telescope (RT) observations of
the Sunyaev-Zel'dovich (SZ) decrement from the cluster Abell 773. The
field contains a number of faint radio sources that required careful
subtraction. We use {\sl ASCA} observations to measure the gas
temperature and a {\sl ROSAT} HRI image to model the gas density
distribution. Normalizing the gas distribution to fit the RT
visibilities returns a value of $H_0$ of $77^{+19}_{-15}\, \rm km \,
s^{-1}\, Mpc^{-1}$ (1-$\sigma$ errors) for an Einstein-de-Sitter
universe, or $85^{+20}_{-17}\, \rm km \, s^{-1}\, Mpc^{-1}$ for a flat
model with $\Omega_{\Lambda} = 0.7$. The errors quoted include
estimates of the effects of the principal errors: noise in the SZ
measurement, gas temperature uncertainty, and line-of-sight depth
uncertainty.
\end{abstract}

\section{Introduction}

We have previously reported the detection of a Sunyaev-Zel'dovich (SZ)
decrement \cite{sandz} towards the $z = 0.217$ cluster Abell~773 using
the Ryle Telescope~(RT)~\cite{grainge93}. 
(The SZ effect in this cluster has also been mapped
by the millimeter array of the Owens Valley
Radio Observatory~\cite{carl96}.)
The RT observations of Abell~773 form part
of a continuing programme to observe an X-ray luminosity-limited
sample of rich, intermediate-redshift clusters in order to measure $
H_0$ by combining SZ and X-ray observations~\cite{jones01}. Such
programmes~(e.g. Reese et al., 2002; Mason, Myers and Readhead, 2001; see
also Birkinshaw, 1999 for a review) 
are direct measurements of $H_0$ free from distance-ladder
arguments.

In Grainge et al.~1993 we did not calculate an estimate of $H_0$ because 
no suitable X-ray image of A773 and no estimate of its gas
temperature existed. A {\sl ROSAT} HRI image and {\sl ASCA} spectroscopic data
have since become available, and we have also made additional RT
observations. These now enable us to make an estimate of the Hubble
constant from this cluster, which, when combined with other clusters
from the sample, will give an estimate of $H_0$ unbiased by the
individual shapes and orientations of the clusters.

\section{Ryle Telescope observations and source subtraction}

The RT \cite{jones91} is an east--west synthesis telescope of 13-m
antennas with a bandwidth of 350~MHz and an average system temperature
for these observations of 65~K at an observing frequency of
15.4~GHz. We used five antennas in a compact configuration, giving two
baselines of 18~m, three of 36~m, and five more out to 108~m. The
short baselines alone are sensitive to the SZ signal; the longer ones
are used to recognize and subtract the radio sources in the field that
would otherwise mask the SZ decrement. We have made a total of 30 12-h
observations of A773, each with the pointing centre $ \rm {RA} \ 09^h
\ 17^m \ 51^s.91 ,\ \rm {Dec.} $ $ +51^{\circ} \ 43' \ 32''
$~(J2000). Phase calibration using 0859+470 and flux calibration
using 3C~48 and 3C~286 were carried at as described in Grainge et al
\shortcite{grainge93}. Similarly, we used the {\sc Postmortem} package
\cite{titterington91} to flag the data for interference and antenna
pointing errors, and to weight them in accord with the continuously
monitored system temperature of each antenna. As a standard check, we
used the {\sc Aips} package to make a map of each 12-h run and then
combined the data.

\begin{figure}
\epsfig{figure=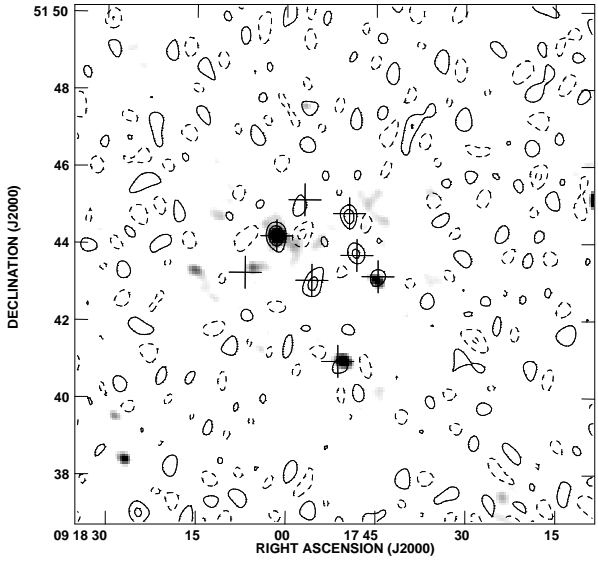, angle = -90, width=80mm}
\caption{Map of the longer baseline-RT data ($2\rm k \lambda$ upwards)
(contours) plus a VLA 1.4~GHz observation (greyscale), jointly used to
determine initial source positions for the source subtraction
procedure. The crosses show the positions of the subtracted sources
listed in Table \ref{table:sources}. The contour interval is $70 \,
\mu$Jy, dashed contours are negative; the greyscale runs from $0.3$
(light) to $1.0$ mJy (dark). The RT data have not been corrected for
the effect of the telescope primary beam (FWHM = 6').
}
\label{vla}
\end{figure}

\begin{figure}
\epsfig{figure=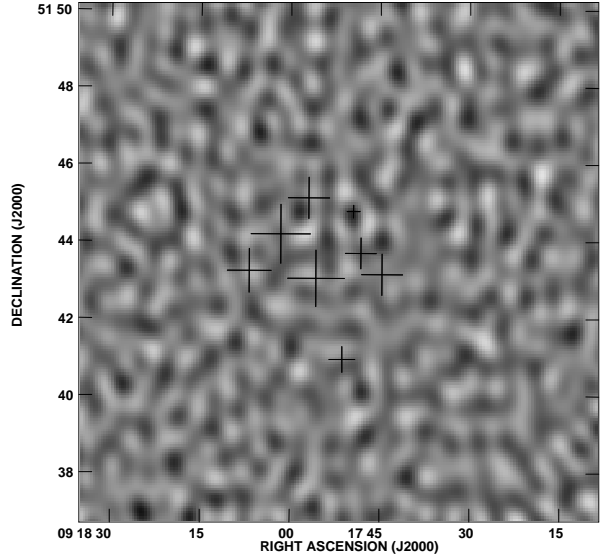, angle = -90, width=80mm}
\caption{Map of the longer baseline data ($2\rm k \lambda$ upwards)
after source subtraction, showing no significant residual
features. The crosses show the positions of the subtracted sources
listed in Table \ref{table:sources}; the size of the cross is
proportional to the flux subtracted. The greyscale runs from $-200$
(light) to $200 \, \mu$Jy (dark). The data have not been corrected for
the effect of the telescope primary beam.}
\label{source-sub}
\end{figure}

We removed radio sources from the data by a simultaneous
maximum-likelihood fit to several point sources and the SZ effect
using a technique described by Grainger et al \shortcite{A611}. We use
a model for the SZ signal as a function of baseline that is based on
the $\beta$-model fit to the X-ray image described below (Section
\ref{xrays}). We simultaneously fit flux densities for trial sources
whose initial positions are determinied both from a map made from just
the long-baseline data ($>2 \, \rm k \lambda$), and from a VLA 1.4-GHz
image of the cluster field (Figure \ref{vla}). This allows us to fit
the optimum flux densities of sources whose existence we know of from
the VLA image but which would not give a significant detection from
the RT data alone. The postitions and fitted flux densities are given
in Table \ref{table:sources}. The image made from the long ($>2\rm k
\lambda$) source-subtracted baselines is consistent with noise (Figure
\ref{source-sub}).

\begin{figure}
\epsfig{figure = 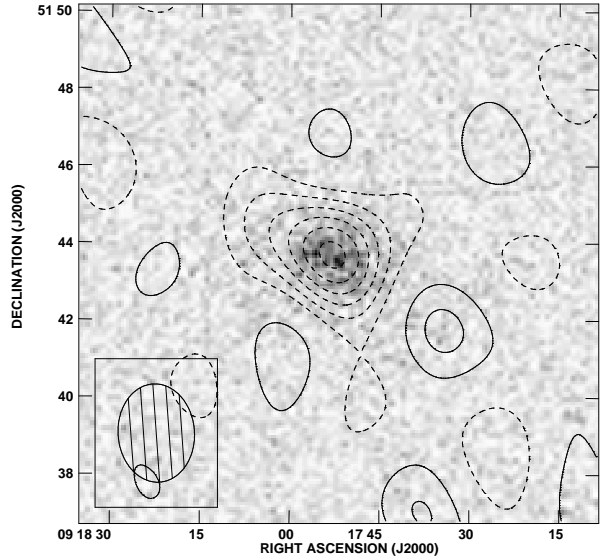,angle=270,width=80mm}
\caption{X-ray and SZ images of A773. The greyscale is the {\sl ROSAT}
HRI image; contours are the source-subtracted and CLEANed RT
image. Contours are in multiples of $80 \, \mu \rm Jy \, beam^{-1}$,
dashed contours are negative. The restoring beam (shown) is $152
\times 119$ arcsec, PA $4^{\circ}$ and the data have not been corrected for
the effect of the telescope primary beam.}
\label{sz+xray}
\end{figure}

To image the decrement, we removed the sources in
Table~\ref{table:sources} from all the visibilities and made a
short-baseline map from baselines shorter than 1 k$\lambda$, and
CLEANed this. The resulting image is shown in
Figure~\ref{sz+xray}. The decrement is of $-527 \, \mu \rm \, Jy\,
beam^{-1}$ with a noise (1-$\sigma$) of $60 \, \mu \rm Jy \,
beam^{-1}$; the beam is $152 \times 119$ arcsec FWHM. Also shown is
the X-ray image of the cluster; it can be seen that the alignment with
the X-ray image is very good. The extension of the SZ image to the
north-east is of marginal significance. The magnitude of the decrement
is consistent with that of $-590 \pm 116 \ \mu$Jy, in the same beam,
reported in Grainge et al \shortcite{grainge93}.

\begin{figure}
\epsfig{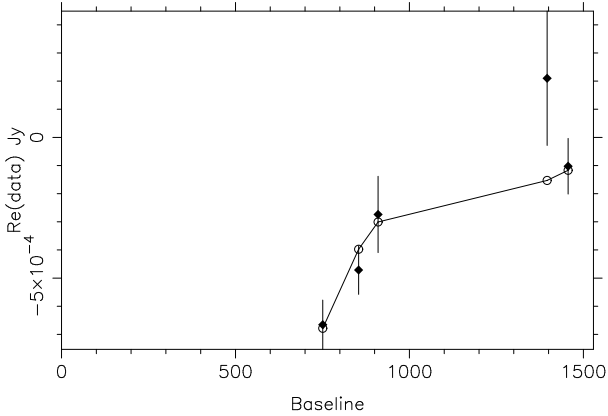}
\caption{The real part of the source-subtracted visibilities as a
function of baseline in wavelengths (filled points with error bars),
with the best-fitting SZ model (open points joined by lines).}
\label{visplot}
\end{figure}

An alternative way of looking at the data is shown in
Figure~\ref{visplot}, which shows the real part of the
source-subtracted visibilities binned radially, along with the
best-fitting model based on the X-ray data. These data have the
advantage, unlike the image pixels, of having independent gaussian
noise on each point; it is these that are used in the fitting for
$H_0$.

\section{X-ray observations and fitting}\label{xrays}

We measure the gas temperature from {\sl ASCA} observations on 
1994 April 29 of 46240 s (GIS) and 39904 s (SIS), using standard {\sc
XSPEC} tools. Times of high background flux were excluded and both GIS
and SIS data were used. We took the Galactic absorbing column density
predicted by Dickey and Lockman \shortcite{dickey90} in the direction
of A773 of $1.3 \times 10^{24}$ H atoms m$^{-2}$. Using a
Raymond-Smith model, we find a temperature of $8.7 \pm 0.7$ keV
($90\%$-confidence error bounds) and a metallicity of 0.25 solar. The
2--10 keV flux from A773 is $6.7{\pm 1.0}$ x $10^{-13}$ W
m$^{-2}$. Our temperature estimate is consistent with that of Allen
and Fabian \shortcite{allen98} who find a temperature of
$9.29^{+0.69}_{-0.60}$ keV ($90\%$-confidence error bounds).

For the X-ray surface-brightness fitting we used a {\sl ROSAT} HRI
image of A773 with an effective exposure of 16518~s obtained on 13--15
April 1994 and analysed using standard {\sc ASTERIX} routines. We
calculate the {\sl ROSAT} HRI count rate, given our estimates of
metallicity and Galactic column and with the K-correction appropriate
to the redshift of A773, to be $1.53 (\pm 0.08) \times 10^{-69}$
counts s$^{-1}$ from a $1 \, \rm m^3$ cube of gas of electron density
$1 \, \rm m^{-3}$ at the temperature of A773 and at a luminosity
distance of 1 Mpc.

We then fitted an ellipsoidal King profile to the X-ray image.  Since
the high spatial resolution of the HRI leads to a low count rate per
pixel, we use Poisson rather than Gaussian statistics to fit for the
measured count in each pixel. For $c_i$ counts measured at position
$x_i$, and for a mean number $f(x_i|a)$ of counts predicted by the
model given parameters $a$ (such as core radius), the probability of
obtaining $c_i$ counts is
\[
P(c_i|a)=\frac{\left(f(x_i|a)\right)^{c_i}}{c_i!}e^{-f(x_i|a)} \ , 
\]
and the most likely value of $a$ can be obtained in a computationally
efficient way by maximizing

\[
\ln P(c|a)=\sum_i (c_i\ln f(x_i|a)-\ln c_i!-f(x_i|a)).
\]

\begin{figure*}
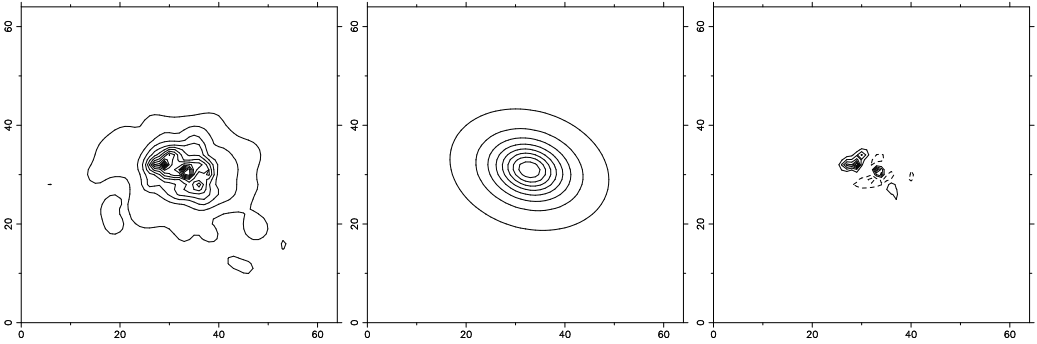

\hbox{
 \epsfig{figure = a773-rosat.eps, angle = -90, width=45mm}
 \epsfig{figure = a773-model.eps, angle = -90, width=45mm}
 \epsfig{figure = a773resid.eps, angle = -90, width=45mm}
}
\caption{X-ray model fits. (left) {\sl ROSAT} HRI image of
A773. (centre) The best-fitting model X-ray surface
brightness. (right) The difference between the data and the model,
which is consistent with the Poisson noise in the data. The contour
interval is 1 count per $8''$ pixel.}
\label{residual}
\end{figure*}

We fitted an ellipsoidal King profile to the HRI data with $\theta_1$
and $\theta_2$ as the perpendicular angular sizes in the plane of the
image, assuming that the length along the line of sight is the
geometric mean of the other two. We find $\theta_1 = 60''$ and
$\theta_2 = 44''$, with the major axis at position angle $16^{\circ}$,
$\beta =0.64$, and central electron density $n_0 = 6.80 \times 10^3 \,
h_{50}^{-1/2} \rm m^{-3}$ where $H_0 = 50\,h_{50}$ km s$^{-1}$
Mpc$^{-1}$. Fig \ref{residual} shows the HRI image, the model, and the
residual image with the best model subtracted. To assess the goodness
of fit, we made 50 realisations of the image with the appropriate
Poisson noise added, and calculated the mean and standard deviation of
their Poisson likelihoods. The likelihood of the observed HRI image is
0.32 standard deviations from the mean; we therefore conclude that the
fit is good and the cluster is well represented by a $\beta$ model.

\begin{figure}
\epsfig{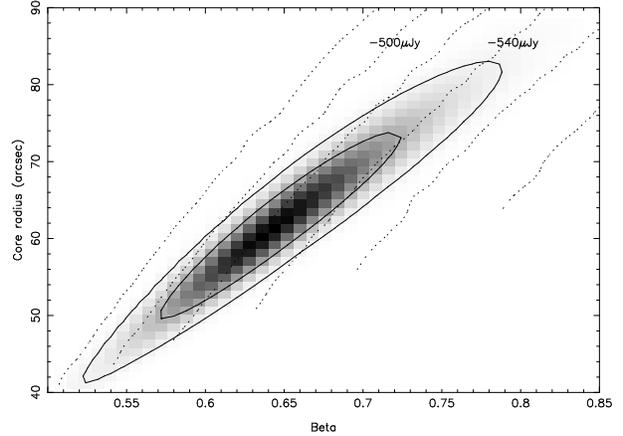}
\caption{Solid contours and greyscale: the likelihood of the fitted
X-ray model as a function of $\beta$ and $\theta_1$, marginalised over
$n_0$, with the axial ratio and position angle fixed at their most
likely values. The solid contours enclose 67\% and 95\% of the
likelihood. Overlaid are dotted contours of the predicted mean flux on
the shortest RT baseline for the given model parameters (using the
most likely value of $n_0$); the contour interval is $40 \mu \rm
Jy$. Since the normalization of the SZ flux relative to the X-ray is
what controls the fit for $h$, this Figure shows the insensitivity of
our $H_0$ determination to the degeneracy in the X-ray model fit.}
\label{margplot}
\end{figure}

There is a strong degeneracy in the fit between $\beta$ and
$\theta_{1,2}$; however this has little effect on the comparison with
the SZ data and the derived value of $H_0$. Figure \ref{margplot}
shows the likelihood contours for the fit in the $\beta$--$\theta_1$
plane, marginalised over $n_0$ and using the best-fit value of the
axial ratio (which is very well constrained). Overlaid are the
contours of predicted mean observed SZ flux density on the shortest RT
baseline. It can be seen that despite the degeneracy between $\beta$
and $\theta_1$, the range of SZ flux densities corresponding to the
1-$\sigma$ limits of the model fit is only $\pm 3 \%$. Since the SZ
flux density varies as $H_0^{-2}$, this corresponds to a $6 \%$ error in
$H_0$ due to the model fitting. This lack of sensitivity to the
$\beta$--$\theta$ degeneracy is characteristic of observations that
are sensitive to spatial frequencies around the cluster core size (see
eg Reese et al \shortcite{reese}) and contrasts with the sensitivity
to the model fitting of measurements that measure only lower spatial
frequencies (eg Birkinshaw \& Hughes\shortcite{birk94}).

\section{$H_0$ estimation}

To measure $H_0$, we compared the real SZ data with a simulation of
the SZ effect from the X-ray gas model. We use the expression of
Challinor \& Lasenby \shortcite{challinor98} to provide a relativistic
correction to the standard non-relativistic SZ expression; in the case
of A773, the effect is to increase our estimate of the $y$-parameter
by $2.4\%$. We then simulated RT observations of the SZ effect due to
the model gas distribution and to compared these with the real
source-subtracted RT visibilities on the same baselines, and adjusted
$H_0$ to get the best fit. Using our temperature of $8.7{\pm 0.7}$ keV
we find $H_0 = 77^{+13}_{-11}$ km s$^{-1}$ Mpc$^{-1}$, assuming an
Einstein-de-Sitter universe. The $1$-${\sigma}$ error quoted is that
due solely to noise in the SZ data. 
For the best fit $\beta, \ \theta_{1,2}$ model,
the corresponding central density
$n_0$ is $8.44 \times 10^3$ m$^{-3}$ and the central decrement $737
\pm 85 {\mu}$K.

Grainge et al \shortcite{grainge02} consider at some length the
contributions to error in the $H_0$ determination from A1413. The
situation in A773 is very similar. The dominant contributions to the
error in $H_0$ in A773 are $\pm 16 \%$ from noise in the SZ
measurement, $\pm 12 \%$ from our estimation of the gas temperature
and a likely error of $\pm 14 \%$ from the uncertain line-of-sight
depth. This is obtained by considering the range of axial ratios of
simulated clusters that is needed to reproduce the projected axial
ratio distribution observed in clusters with redshift similat to that
of A773 \cite{will_thesis}. Clearly this estimate is rather uncertain
for a single object, but can be significantly reduced by averaging a
sample of clusters with random orientations. Table \ref{errors} shows
the complete error budget, and the final 1-$\sigma$ error limits of
$H_0 = 77^{+19}_{-15}\, \rm km \, s^{-1}\, Mpc^{-1}$ if $(\Omega_{\rm
m}, \Omega_{\Lambda}) = (1.0, 0.0)$ and $H_0 = 85^{+20}_{-17}\rm \, km
\, s^{-1}\, Mpc^{-1}$ if $(\Omega_{\rm m}, \Omega_{\Lambda}) = (0.3,
0.7)$.

\section{Conclusions}

Using {\sl ASCA}, {\sl ROSAT} HRI, and RT observations of A773, we
find:
\begin{enumerate}

\item there are eight radio sources detectable in the field of the
cluster that we have removed from the data, which would otherwise
contaminate the measurement of the SZ effect;

\item the correlated fitting errors on the shape parameters $\beta$
and $\theta$ have negligable effect on the derived value of $H_0$, a
feature characteristic of observations on the scale of the cluster
core size;

\item the estimated value of $H_0$ is $77^{+19}_{-15}\, \rm km \,
s^{-1}\, Mpc^{-1}$ if $(\Omega_{\rm m}, \Omega_{\Lambda}) = (1.0,
0.0)$ or $85^{+20}_{-17}\, \rm km \, s^{-1}\, Mpc^{-1}$ if
$(\Omega_{\rm m}, \Omega_{\Lambda}) = (0.3, 0.7)$ , where the
1-$\sigma$ error bars include estimates from the main sources of
error---noise in the SZ data, X-ray temperature uncertainty, and
uncertain line-of-sight depth.

\end{enumerate}

\subsection*{ACKNOWLEDGMENTS}

We thank the staff of the Cavendish Astrophysics group who maintain
and operate the Ryle Telescope, which is funded by PPARC. 
AE acknowledges
support from the Royal Society; WFG acknowledges the support of a
PPARC studentship; RK acknowledges support from an EU Marie Curie
Fellowship.


This paper has been produced using the Blackwell Scientific Publications \LaTeX
\ style file. 

\clearpage

\begin{table}
\begin{tabular}{ccc}
Flux density /$\mu$Jy & RA offset /$^{\prime \prime}$ & dec offset /$^{\prime \prime}$\\
\hline
163 & $-67$ & $-23$\\
123 & $-34$ & $9$\\
55  & $-22$ & $75$\\
103 & $-5$ & $-155$\\
228 & $35$ & $-29$\\
162 & $46$ & $96$\\
238 & $90$ & $40$\\
177 & $139$ & $-16$\\
\end{tabular}
\caption{Flux densities and positions of sources removed from the A773
field.  Flux densities are apparent, ie not corrected for the RT
primary beam response, and are all $\pm 35\, \mu$Jy. Offsets are
relative to the pointing centre of $\rm 09^h 17^m 51^s.91 ~
+51^{\circ} 43' 32''$ (J2000).}
\label{table:sources}
\end{table}

\begin{table*}
\begin{tabular}{ll}
Source of error & Error contribution to $H_0$\\
\hline
SZ measurement & $\pm 16\%$\\
Line-of-sight depth & $\pm 14\%$\\
Gas temperature & $\pm 12\%$\\
Error in calculated X-ray emission constant & $\pm 6\%$\\
Primary flux calibration & $\pm 2.5\% $\\
\hline
Total (in quadrature)& $\pm 25\%$ \\
\end{tabular}
\caption{Error budget for $H_0$ calculation; the dominant terms are
the errors in the SZ measurement and the gas temperature. The errors
are all multiplicative, ie the 1-$\sigma$ limits on $H_0$ are $77
\times 1.25 = 96$ and $77 / 1.25 = 62 \, \rm km\,s^{-1}\,Mpc^{-1}$.}
\label{errors}
\end{table*} 

\begin{thebibliography}{99}
\bibitem[\protect\citename{Allen \& Fabian }1998]{allen98}Allen S.W., Fabian
A.C., 1998, MNRAS, 297, L57
\bibitem[\protect\citename{Birkinshaw \& Hughes
}1994]{birk94}Birkinshaw M., Hughes J.P., ApJ, 1994, 420, 33
\bibitem[\protect\citename{Birkinshaw
}1999]{birk99}Birkinshaw M., Phys.Rept. 310, 97
\bibitem[\protect\citename{Carlstrom, Joy \& Grego
}1996]{carl96}Carlstrom, J.E., Joy, M., Grego, L., ApJ, 456, L75 and
ApJ, 461, L59
\bibitem[\protect\citename{Challinor \& Lasenby }1998]{challinor98}Challinor A.,
Lasenby A., ApJ, 1998, 499, 1
\bibitem[\protect\citename{Dickey \& Lockman }1990]{dickey90}Dickey J.M., Lockman F.J., 1990,
Ann. Rev.  Astron. Astrophys., 28, 215
\bibitem[\protect\citename{Grainge et al }1993]{grainge93}Grainge, K., Jones,
M., Pooley, G.G., Saunders, R., Edge, A.C., 1993, MNRAS, 265, L57
\bibitem[\protect\citename{Grainge et al }2002]{grainge02}Grainge, K., Jones,
M.E., Pooley, G.G., Saunders, R., Edge, A.C., Kneissl, R., 2002, MNRAS
333, 318
\bibitem[\protect\citename{Grainger et al }2002]{A611}Grainger, W. F.,
Das, R.,
Grainge, K., Jones, M.E., Kneissl, R., Pooley, G.G., Saunders, R.,
 2002, MNRAS, 337, 1207
\bibitem[\protect\citename{Grainger }2001]{will_thesis}Grainger, W.F., PhD thesis, University of Cambridge, 2001
\bibitem[\protect\citename{Jones }1991]{jones91}Jones M.E., 1991, in Cornwell
T.J., Perley R., eds, Proc. IAU Colloq. 131, ASP Conf. Ser. 19, Radio
Interferometry: Theory, Techniques and Applications. Astron. Soc. Pac., San
Francisco, p. 395
\bibitem[\protect\citename{Jones et al.}2001]{jones01} Jones M.E. et
al., 2001, submitted to MNRAS
\bibitem[\protect\citename{Mason, Myers \& Readhead }2001]{mason01}
Mason, B.S., Myers, S.T., Readhead, A.C.S., 2001, ApJ, 555, L11
\bibitem[\protect\citename{Reese et al }2000]{reese}Reese E.D, Mohr
J.J., Carlstrom J.E., Joy M., Grego L., Holder G.P., Holzapfel W.L.,
Hughes J.P., Patel S.K., Donahue M., 2000, ApJ, 533, 38
\bibitem[\protect\citename{Reese et al }2002]{reese02}Reese E.D,
Carlstrom, J.E., Joy, M., Mohr, J.J., Grego, L., Holzapfel W.L., 2002,
submitted to ApJ
\bibitem[\protect\citename{Sunyaev \& Zel'dovich} 1972]{sandz}Sunyaev, R. A.,
Zel'dovich, Ya B., 1972, Comm. Astrophys. Sp. Phys., 4, 173
\bibitem[\protect\citename{Titterington }1991]{titterington91}Titterington
D.J., 1991, in Cornwell T.J., Perley R., eds, Proc. IAU Colloq. 131, ASP
Conf. Ser. 19, Radio Interferometry: Theory, Techniques and
Applications. Astron. Soc. Pac., San Francisco, p. 128

\end{thebibliography}
\end{document}